\documentclass[prb,twocolumn,showpacs]{revtex4}
\usepackage{amssymb}

\usepackage{graphicx}

\usepackage[T1]{fontenc}

\usepackage[american]{babel}

\begin{document}

\title{Thickness-dependence of the electronic properties in V$_2$O$_3$ thin films}

\author{C.~Grygiel}
\affiliation{Laboratoire CRISMAT, UMR 6508 CNRS-ENSICAEN, 6, Boulevard
             du Mar\'echal Juin, 14050 CAEN Cedex, France} 
\author{Ch.~Simon}
\affiliation{Laboratoire CRISMAT, UMR 6508 CNRS-ENSICAEN, 6, Boulevard
             du Mar\'echal Juin, 14050 CAEN Cedex, France} 
\author{B.~Mercey}
\affiliation{Laboratoire CRISMAT, UMR 6508 CNRS-ENSICAEN, 6, Boulevard
             du Mar\'echal Juin, 14050 CAEN Cedex, France} 
\author{W.~Prellier}
\affiliation{Laboratoire CRISMAT, UMR 6508 CNRS-ENSICAEN, 6, Boulevard
             du Mar\'echal Juin, 14050 CAEN Cedex, France} 
\author{P.~Limelette}
\affiliation{Laboratoire LEMA, CNRS-CEA UMR 6157, Université F.~Rabelais, 
                        Parc de Grandmont, 37200 TOURS, France} 
\author{R.~Fr\'esard}
\affiliation{Laboratoire CRISMAT, UMR 6508 CNRS-ENSICAEN, 6, Boulevard
             du Mar\'echal Juin, 14050 CAEN Cedex, France}

\date{\today}

\vspace{8.5cm}

\begin{abstract}

High quality vanadium sesquioxide V$_2$O$_3$ films (170-1100\,{\AA}) were grown using the pulsed
laser deposition technique on (0001)-oriented sapphire substrates, and the effects
of film thickness on the lattice strain and electronic properties were
examined. X-ray diffraction indicates that there is an in-plane
compressive lattice parameter ($a$), "close to -3.5\% with respect to the substrate" and an out-of-plane tensile lattice
parameter ($c$) . 
The thin film samples display metallic character between 2-300\,{K}, and no metal-to-insulator
transition is observed. At low temperature, the V$_2$O$_3$ films behave as a strongly
correlated metal, and the resistivity ($\rho $) follows the equation $\rho
$=$\rho _0+A\cdot T^2$, where $A$ is the transport coefficient in a Fermi
liquid. Typical values of $A$ have been calculated to be 0.14\,{$\mu\Omega$  cm K$^{-2}$}, which is 
in agreement with the coefficient reported for V$_2$O$_3$ single crystals
under high pressure. Moreover, a strong temperature-dependence 
of the Hall resistance confirms the electronic correlations of these V$_2$O$_3$ thin films samples.

\end{abstract}


\maketitle

\newpage
\quad 
First discovered by Foex in 1946,\cite{foex46}  vanadium sesquioxide
V$_2$O$_3$ has received a great deal of attention, both by theoreticians
as well as experimentalists. Indeed, it has been recognized previously that a pressure-induced metal-to-insulator transition (MIT) for V$_2$O$_3$ is driven by
electron correlation,
\cite{mott49} establishing V$_2$O$_3$ as a prototypical strongly correlated
electron system. As a result, numerous studies on the effect of composition or
external parameters on the transport properties of V$_2$O$_3$ have been
reported.\cite{whan73,yethiraj90,limelette2003} 
Particular attention has been also paid to the phase transitions of V$_2$O$_3$: in the
pressure-temperature plane two phase transitions are reported, either when
applying 
a hydrostatic pressure or a chemical pressure 
(see for example, (V$_{1-x}$M$_x$)$_2$O$_3$ with M=Cr, Ti,..).\cite{whan73,yethiraj90}
For example, a system that is close to all phase boundaries is (V$_{0.985}$Cr$_{0.015}$)$_2$O$_3$ at
200\,K.\cite{kuwamoto80} When the temperature is decreased, this paramagnetic metal undergoes
a first order phase transition from a corundum structure with rhombohedral
symmetry ($R\overline{3}c$ space group, with
\textit{a}=4.951\,{\AA}, \textit{c}=14.003\,{\AA}) to an antiferromagnetic
insulator 
with monoclinic structure ($I2/a$).\cite{dernier70} When the temperature or 
the Cr-content is increased, a paramagnetic metal to paramagnetic insulator
transition takes place.\cite{kuwamoto80} While the former transition bears
strong similarities with many usual magnetic transitions, the latter one
corresponds to the famous Mott transition. 
Qualitatively, this Mott transition is well described by the Hubbard model:
increasing the Cr-content results in increasing the ratio $U/W$ ($U$ being the
strength of the Coulomb interaction, and $W$ the bandwidth), in which case the
quasiparticle residue decreases, and eventually vanishes at the
MIT.\cite{Brinckman} Furthermore, when considering the optical conductivity,
the optical weight at low energy is transfered to energies of order
$U$.\cite{Thomas} When the system is close to the Mott point, thermal
fluctuations destabilize the coherence of the Fermi liquid (FL), and an increase
in temperature results into a MIT.\cite{Spalek,georges96,fresard97} Several orbitals are involved at the Fermi energy, and however, a quantitative
description of V$_2$O$_3$ needs to build on a more involved theory, such as
the one pioneered by Held {\em et al}.\cite{Held}

Despite extensive studies on polycrystalline powder and single crystal
V$_2$O$_3$ samples,\cite{whan70,ueda80,shivashankar83,whan69} there have been few
reports on thin film samples. Schuler \textit{et al.} addressed the influence of the
synthesis conditions upon the classical metal-to-insulator transition and the
growth modes of the films.\cite{schuler97} The relation between the
transition temperatures and the lattice parameters also have been 
reported.\cite{autier2006,yonezawa2004,luo2004} There is little knowledge however on the
thickness-dependence of the properties of the V$_2$O$_3$ thin films. 

In this letter, we examine a series of epitaxial V$_2$O$_3$ thin films, including 
their lattice parameters,
roughness, and Hall resistance, in order to investigate their transport properties as a function of thickness.

Samples of V$_2$O$_3$ thin films, of which the thickness ranged from 170\,{\AA} to 1100\,{\AA},
were grown on (0001)-oriented Al$_2$O$_3$ substrates (rhombohedral with the parameters, \textit{a}=4.758\,{\AA}, \textit{c}=12.991\,{\AA}) 
by the pulsed laser deposition technique. A pulsed KrF excimer laser (Lambda Physik, Compex, $\lambda$=248\,{nm}) was focused onto 
a stoichiometric V$_2$O$_5$ target at a fluence of 4\,{J cm$^{-2}$} with a repetition rate of 3\,{Hz}. The substrate
heater was kept at a constant temperature ranging from 600${{}^{\circ }}$C to 650${{}^{\circ }}$C. 
A background of argon pressure around 0.02\,{mbar} was applied inside the chamber. At the end of the deposition, the film was cooled down 
to room temperature at a rate of 
10\,K min$^{-1}$ 
under a 0.02\,{mbar} argon pressure. 
The film thickness was determined, to an uncertainty below two V$_2$O$_3$ unit cells, by a mechanical stylus measuring system (Dektak$^{3}$ST). 
The structure of the films was examined by X-Ray Diffraction (XRD, with the Cu K$\alpha_1$ radiation, $\lambda $=1.54056\,{\AA}) using a Seifert 3000P
diffractometer for the out-of-plane measurements and a Philips X'Pert
for the in-plane measurements. In-plane \textit {a} lattice parameters were extracted from (\textit {104}), (\textit{116}), (\textit{113}) asymmetric reflections. 
The resistivity of the samples were measured in four probe configuration using a PPMS system. To make appropriate connections onto the
film, four silver plots were first deposited via thermal evaporation onto
the film, and then thin aluminum contact wires were used to connect these
areas to the electrodes. For Hall effect measurements, a Van der Pauw
configuration of silver plots was used.
For each temperature value, the transverse resistance is measured with an
applied magnetic field varying from -7\,T up to +7\,T. The Hall resistance R$_H$ is calculated from the transverse resistance ($R_{xy}$) using the formula R$_H$=$(t\cdot R_{xy})/H$, 
where $H$ is the applied magnetic field.

As it is known
that the nonstoichiometry can drastically influence the 
electric properties of V$_2$O$_3$,\cite{ueda80} a X-ray photoelectron spectroscopy study
was carried out. It shows
that the oxidation state of vanadium can be estimated around $+3$,
confirming that the films stoichiometry is close to V$_2$O$_3$.\cite{autier2006}
$\Theta-2\Theta$ scans XRD measurements reveal that only the peaks corresponding to
the \textit{00l} reflections (where $l$=6, 12...) are present, which indicates that
the $c$-axis of
the films is perpendicular to the plane of the substrate. 
$\Phi $-scans, recorded around the (\textit{104})
reflection show three peaks separated by 120${{}^{\circ }}$ from each other,
indicating that the films have a 3-fold symmetry and are grown epitaxially, with respect to the substrate.  
These results are in agreement with the rhombohedral symmetry observed  in the
bulk V$_2$O$_3$. This symmetry and the $R\overline{3}c$ space group are further confirmed by electron
diffraction pattern analyses. The high quality of the films was also attested by the low value of the
rocking-curve close to (0.20${{}^{\circ }}$) measured around the (\textit{006})
reflection of the film.
\begin{figure}[h!]
\begin{center}
\includegraphics*[width=9.2cm]{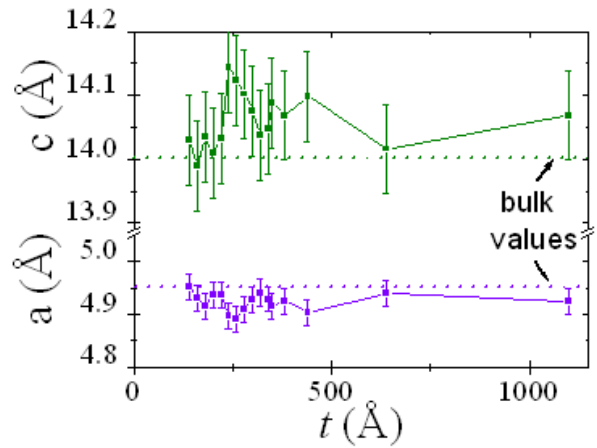}
\end{center}
\caption{Thickness-dependence of the \textit{a, c} lattice parameters. Doted lines indicate the bulk values. The line is a guide for the eye only.
}
\label{fig:F_1}
\end{figure}

Fig.~\ref{fig:F_1} displays  the evolution of the lattice parameters ($c$) and ($a$),
characteristic of the $R\overline{3}c$ structure, as a
function of the thickness $t$. 
The out-of-plane lattice parameter ($c$) is
slightly larger than the bulk values, by 0.45$\pm$0.1 \%, while the in-plane ($a$) is
smaller by -0.55$\pm$0.1 \%,
confirming a biaxial compression in the (\textit{ab}) plane, estimated to
-3.5$\pm$0.5 \% with respect to the substrate (the stress values are calculated from the mean-values ($c$) and ($a$) obtained from the thickness dependence in Fig.~\ref{fig:F_1}). This suggests an anisotropic strain
similarly to previous reports.\cite{yonezawa2004} Surprisingly, the 
lattice parameters are almost independent of $t$, indicating
that the films are fully strained in the whole thickness 
range. 
The surface morphology was also studied by Atomic Force Microscopy
(AFM) with a scan area of $3\times3\mu$m$^{2}$. 
Topography of the V$_2$O$_3$ samples reveals that the roughness
increases when the thickness increases. For example, a typical surface
roughness 
(rms) of 4.5\,{\AA}\  was observed for the thinnest film (170\,{\AA}), 
while a thicker sample (700\,{\AA}) has a rms value of
11\,{\AA}. This may indicate that the growth mode is mixed: a layer-by-layer
(2D-mode) on the ({\textit ab}) plane at the 
initial step of the growth, and an island coalescence (3D-mode) along the
\textit{c}-axis when the thickness increases. To summarize, the structural and
microstructural analyses confirm that the films crystallize in the 
$R\overline{3}c$ structure, as in bulk, despite a large in-plane
compressive strain. 

\begin{figure}[h!]
\begin{center}
\includegraphics*[width=9.2cm]{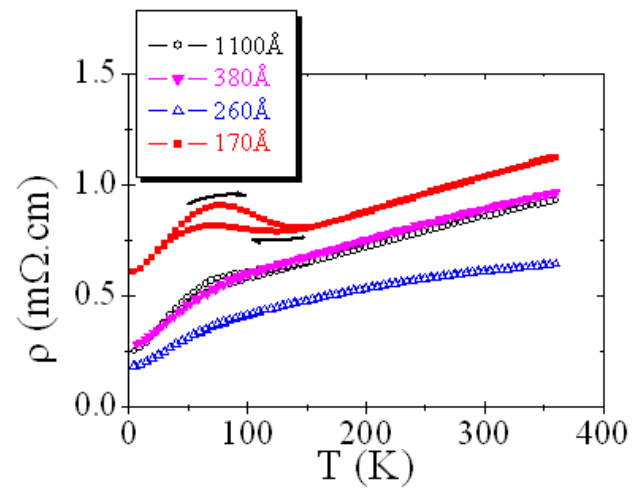}
\end{center}
\caption{Temperature-dependence of the longitudinal resistivity of several V$_2$O$_3$ thin
films.
}
\label{fig:F_2}
\end{figure}

\begin{figure}[h!]
\begin{center}
\includegraphics*[width=10.2cm]{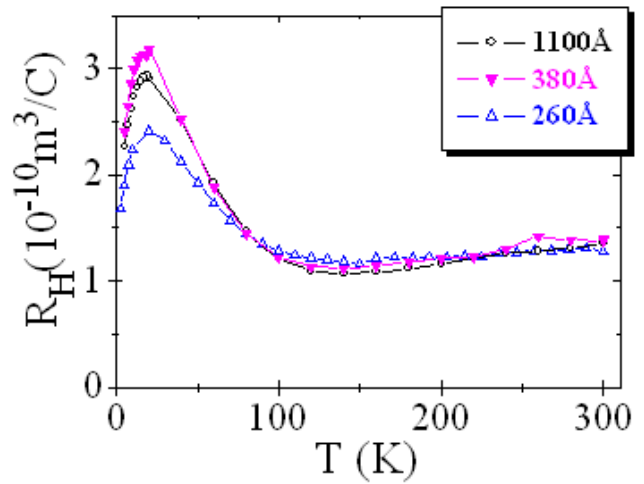}
\end{center}
\caption{Temperature-dependence of the Hall resistance of several V$_2$O$_3$ thin films.
}
\label{fig:F_4}
\end{figure}

The longitudinal
resistivity ($\rho$) of V$_2$O$_3$ films is plotted in Fig.~\ref{fig:F_2} as a
function of 
temperature for several thicknesses. In contrast to bulk samples, none of the
investigated films exhibit a strong temperature-dependent resistivity. This
is especially valid for $t>$ 220\,{\AA}, in which case the resistivity
increases continuously with temperature as in a metal. Nevertheless, some
hysteresis is observed, though strongly suppressed with respect to the
bulk. Thus, we can conjecture that the abovefound stress in the ({\textit ab})
plane results into a larger band width, and hinders the paramagnetic metal-to-antiferromagnetic insulator transition. In
contrast, for thinner films (t < 220\,{\AA}), the films are metallic with a weak
increase of the resistivity near $150K$. Its origin might be a reminder of the structural transition.  
\begin{figure}[h!]
\begin{center}
\hspace*{-.5cm}\includegraphics[width=9.7cm]{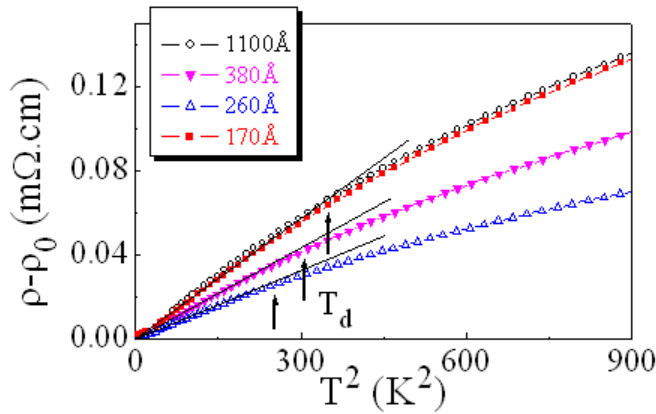}
\end{center}
\caption{Resistivity versus $T^2$ in the low temperature region.  The $T_d$ temperature (see text for details) is also indicated.
}
\label{fig:F_3}
\end{figure}
\begin{table}[b!]
\caption
{
Thickness $t$, residual resistivity $\rho _0$, Fermi liquid transport
coefficient $A/A_0$ (with $A_0 = 10^{-5}$m$\Omega$ cm K$^{-2}$), temperature $T_d$, carriers number at 300\,K $n$, and Hall
mobility at 300\,K $\mu_H $  for three films.
}
\label{tab:1}
\begin{ruledtabular}
\begin{tabular}{cccccc}
\multicolumn{1}{c}{$t$ ({\AA})}                        &
\multicolumn{1}{c}{$\rho _0$ (m$\Omega$ cm)}    &
\multicolumn{1}{c}{$A/A_0$ 
}   &
\multicolumn{1}{c}{$T_d$ (K)}  &
\multicolumn{1}{c}{$n$ (cm$^{-3}$)}                  &
\multicolumn{1}{c}{$\mu_{H} $ (cm$^{2}$ (V\,s)$^{-1}$)}          \\
\colrule
1100 & 0.24 & 19.6 & 18.7& 4.5$\pm0.2 10^{22}$ & 0.16$\pm $0.05 \\
380 & 0.27 & 14.0 & 17.5 & 4.4$\pm0.4 10^{22}$ & 0.15$\pm $0.06 \\
260 & 0.19 & 10.2 & 15.8 & 4.9$\pm0.5 10^{22}$ & 0.21$\pm $0.10\\
\end{tabular}
\end{ruledtabular}
\end{table} 
As shown in Fig.~\ref{fig:F_3}, resistivity data for all films follow a law in $\rho=\rho _0+A\cdot T^2$ in the
temperature range from $2K$ up to a characteristic temperature $T_d$. 
Here $\rho _0$ represents the (film dependent) residual resistivity, 
and $A$ the transport coefficient in a Fermi liquid. This behavior, which differs from simple metals that exhibit $T^3$ or $T^5$ behavior, is seldom
observed over such a temperature range, except for a few strongly correlated
electron system, such as Ca$_3$Co$_4$O$_9$.\cite{limelette2005}
Table~\ref{tab:1} summarizes the values ($\rho_0$), ($A$) and ($T_d$) for a series of films. The
average value of $A$ is close to 0.14\,{$\mu\Omega$ cm K$^{-2}$} across the whole
thickness range, indicating that the $A$ coefficient is not
\textit{significantly} thickness-dependent. Moreover, it should be note that such values are 
similar to those measured for single crystal V$_2$O$_3$ samples subjected to
high pressures (26-52\,kbar).\cite{whan69} The high $A$ values
involve the existence of strong electronic correlations in our metallic thin 
films. The product $A\cdot (T_d)^2/a$ is about two orders of magnitude lower than $h/e^2$, indicating
that the FL behavior does not extend up to the effective Fermi temperature. Instead, an additional scattering channel opens up at $T$ > $T_d$, 
which might be provided by spin fluctuations. To better quantify the temperature range where this scattering channel is relevant, 
we performed Hall effect measurements in the temperature range 2-300\,{K}. 
Fig.~\ref{fig:F_4} shows the resulting Hall resistance ($R_H$) for a series
of films. 
 
The positive slope of the Hall resistance implies 
hole-like charge carriers, as inferred by McWhan \textit{et al}.\cite{whan70}
Moreover, while the Hall resistance would be nearly temperature independent in a regular metal,  
it here exhibits a strong temperature dependence, especially for $T$ < 200\,{K}.
In particular, the strain involved in our films has little influence on the location of the maximum of R$_H$, as it is located at a temperature very close 
to the one reported for metallic bulk samples.\cite{carter93} Nevertheless, this temperature slightly increases as the thickness decreases. 
A similar result has been observed in the aforementioned single crystals with an increase in pressure, indicating that the decrease of the thickness is consistent with an increase of the pressure.
 At room temperature, in the temperature-independent regime, the values of the carriers number ($n$) and Hall mobility ($\mu_H$) can also be extracted from the Hall resistance value (see Table I). 
For the thicker film (1100\,{\AA}), they are calculated to be $n=4.5\pm0.2 10^{22}cm^{-3}$ and $\mu_H$=0.16$\pm $0.05\,{cm$^{2}$ V$^{-1}$ s$^{-1}$}, which are consistent
 with the ones observed in V$_2$O$_3$ single crystals.\cite{whan70} 
 Note that $R_H$ looses its temperature-dependence above $\sim$ 200\,{K}, which indicates that above this temperature the correlation length 
of the fluctuations
 that scatter the electrons become of the order of the lattice parameter. 
Consequently, in this regime, the resistivity is expected to be $T$-linear, which is indeed observed for most of the films.

In summary, high quality epitaxial V$_2$O$_3$ thin films were grown, with thickness ranging from 170-1100\,{\AA}, by pulsed laser deposition on sapphire substrate (0001-Al$_2$O$_3$).
Using multiple of characterization techniques, we confirm that the films have the same rhombohedral structure ($R\overline{3}c$ space group) as in the bulk despite the 
substrate-induced strains. The thickness-dependence of the electronic properties show the suppression of the classical metal-to-insulator transition with a metal-like behavior.
At low temperature, the dependence of the resistivity as a function of $T^2$ was measured to be that of a strong correlated metal, and was confirmed 
by temperature-dependence of the Hall resistance.
 
We thank D. Grebille and A. Pautrat for fruitful discussions.  J.F. Hamet and Y. Thimont are also acknowledged for the AFM measurements.
This work is carried out in the frame of the STREP CoMePhS  (NMP3-CT-2005-517039)
supported by the European community and by the CNRS, France. 

\end{document}